\documentclass[aps,prl,twocolumn,superscriptaddress,nofootinbib,floatfix]{revtex4-2}
\pdftrailerid{}
\usepackage[T1]{fontenc}
\usepackage[utf8]{inputenc}
\usepackage{amsmath,amssymb}
\usepackage{graphicx}
\usepackage{booktabs}
\usepackage{xcolor}
\usepackage[colorlinks=true,linkcolor=blue!50!black,citecolor=blue!50!black,urlcolor=blue!50!black]{hyperref}
\hypersetup{pdftitle={},pdfauthor={},pdfsubject={},pdfkeywords={},pdfcreator={},pdfproducer={}}
\graphicspath{{./}}

\newcommand{\hMpc}{\,h^{-1}\,\mathrm{Mpc}}
\newcommand{\zeff}{z_{\rm eff}}
\newcommand{\DV}{D_{V}}
\newcommand{\DM}{D_{M}}
\newcommand{\DHz}{D_{H}}
\newcommand{\rd}{r_{\rm d}}
\newcommand{\apar}{\alpha_{\parallel}}
\newcommand{\aperp}{\alpha_{\perp}}
\newcommand{\aiso}{\alpha_{\rm iso}}
\newcommand{\WH}{WH24}
\newcommand{\heading}[1]{\textit{#1}---\ignorespaces}
\newsavebox{\fitbox}
\newcommand{\fitwidth}[1]{\sbox{\fitbox}{#1}%
  \ifdim\wd\fitbox>\textwidth\resizebox{\textwidth}{!}{\usebox{\fitbox}}\else\usebox{\fitbox}\fi}
\providecommand{\apj}{Astrophys. J.}\providecommand{\apjs}{Astrophys. J. Suppl. Ser.}
\providecommand{\mnras}{Mon. Not. R. Astron. Soc.}
\providecommand{\aap}{Astron. Astrophys.}\providecommand{\jcap}{J. Cosmol. Astropart. Phys.}
\providecommand{\prd}{Phys. Rev. D}

\newcommand{\vNclus}{302\,176}
\newcommand{\vArea}{10\,793}
\newcommand{\vzeffv}{0.405}
\newcommand{\vtrd}{99.60}
\newcommand{\vDVdesi}{10.42}
\newcommand{\vDVplanck}{10.57}
\newcommand{\vnjk}{100}
\newcommand{\vtsperp}{6.72}
\newcommand{\vtspar}{11.59}
\newcommand{\valphapost}{0.974}
\newcommand{\valphapoststat}{0.011}
\newcommand{\vDVpost}{10.29}
\newcommand{\vDVpoststat}{0.12}
\newcommand{\vDVpostsys}{0.09}
\newcommand{\vfracposttot}{1.5}
\newcommand{\vchisqpost}{28.9}
\newcommand{\vptepost}{0.02}
\newcommand{\vdchisqpost}{31.9}
\newcommand{\vsigpost}{5.6}
\newcommand{\vsigposttwo}{5.65}
\newcommand{\vrecgain}{1.6}
\newcommand{\valphapre}{0.980}
\newcommand{\valphaprestat}{0.018}
\newcommand{\vDVpre}{10.36}
\newcommand{\vDVprestat}{0.19}
\newcommand{\vDVpresys}{0.08}
\newcommand{\vfracpretot}{2.0}
\newcommand{\vchisqpre}{6.9}
\newcommand{\vdchisqpre}{12.1}
\newcommand{\vsigpre}{3.5}
\newcommand{\vsigpretwo}{3.48}
\newcommand{\vaperppost}{0.981}
\newcommand{\vaperppoststat}{0.014}
\newcommand{\vaparpost}{0.953}
\newcommand{\vaparpoststat}{0.034}
\newcommand{\vanisocorrpost}{-0.42}
\newcommand{\vDMpost}{10.78}
\newcommand{\vDMpoststat}{0.16}
\newcommand{\vDHpost}{23.03}
\newcommand{\vDHpoststat}{0.82}
\newcommand{\vanisosigpost}{5.9}
\newcommand{\vanisochisqpost}{45.9}
\newcommand{\vaperppre}{0.984}
\newcommand{\vaperpprestat}{0.023}
\newcommand{\vaparpre}{0.961}
\newcommand{\vaparprestat}{0.055}
\newcommand{\vanisocorrpre}{-0.44}
\newcommand{\vanisosigpre}{3.6}
\newcommand{\vanisochisqpre}{19.4}
\newcommand{\vgtwon}{50}
\newcommand{\vdsig}{+0.09}
\newcommand{\vdsigerr}{0.07}
\newcommand{\vpostn}{41}
\newcommand{\vpostda}{-0.0005}
\newcommand{\vpostdaerr}{0.0013}
\newcommand{\vpostdsig}{-0.11}
\newcommand{\vpostdsigerr}{0.11}
\newcommand{\vdampfac}{0.784}
\newcommand{\vdampperp}{5.27}
\newcommand{\vdamppar}{9.09}
\newcommand{\vcovn}{100}
\newcommand{\vcovDpre}{1.115}
\newcommand{\vcovDpost}{0.989}
\newcommand{\vnsdesipost}{0.8}
\newcommand{\vnsplanckpost}{1.8}
\newcommand{\vnsdesipre}{0.3}
\newcommand{\vnsplanckpre}{1.0}
\newcommand{\vratiostat}{2.6}
\newcommand{\vratiotot}{2.0}
\newcommand{\vratiohong}{3.2}
\newcommand{\voffmin}{0.01}
\newcommand{\voffmax}{0.36}
\newcommand{\vsigsysisopre}{0.008}
\newcommand{\vsigsysisopost}{0.009}
\newcommand{\vsigsysperppre}{0.012}
\newcommand{\vsigsysperppost}{0.014}
\newcommand{\vsigsysparpre}{0.017}
\newcommand{\vsigsysparpost}{0.021}
\newcommand{\vDMsyspost}{0.16}
\newcommand{\vDHsyspost}{0.50}
\renewcommand{\vDVpostsys}{0.09}
\renewcommand{\vfracposttot}{1.5}
\renewcommand{\vDVpresys}{0.08}
\renewcommand{\vfracpretot}{2.0}
\renewcommand{\vnsdesipost}{0.8}
\renewcommand{\vnsplanckpost}{1.8}
\renewcommand{\vnsdesipre}{0.3}
\renewcommand{\vnsplanckpre}{1.0}
\renewcommand{\vratiotot}{2.0}
\renewcommand{\vratiohong}{3.2}
\renewcommand{\voffmin}{0.01}
\renewcommand{\voffmax}{0.35}
\newcommand{\vTNgal}{42\,590}
\newcommand{\vTzeff}{0.0475}
\newcommand{\vTnmock}{100}
\newcommand{\vTpreA}{0.949}
\newcommand{\vTpreAup}{0.034}
\newcommand{\vTpreAlo}{0.032}
\newcommand{\vTpriorlo}{0.80}
\newcommand{\vTpriorhi}{1.20}
\newcommand{\vTpresig}{1.2}
\newcommand{\vTpostsig}{0.9}
\newcommand{\vTedgepre}{63}
\newcommand{\vcthreeApost}{0.49}
\newcommand{\vcthreeapost}{1.72}
\newcommand{\vcthreenbrpost}{0.77}
\newcommand{\vcthreenbrjkpost}{0.84}
\newcommand{\vcthreetwentyfivepost}{0.40}
\newcommand{\vcthreexipost}{0.05}
\newcommand{\vcthreegausspost}{1.26}
\newcommand{\vcthreejkmedpost}{1.03}
\newcommand{\vcthreediagjkpost}{1.36}
\newcommand{\vcthreejklopost}{$-$15}
\newcommand{\vcthreejkhipost}{+32}
\newcommand{\vcthreeApre}{0.60}
\newcommand{\vcthreeapre}{1.65}
\newcommand{\vcthreenbrpre}{0.78}
\newcommand{\vcthreenbrjkpre}{0.85}
\newcommand{\vcthreetwentyfivepre}{0.41}
\newcommand{\vcthreexipre}{0.05}
\newcommand{\vcthreegausspre}{1.15}
\newcommand{\vcthreejkmedpre}{1.02}
\newcommand{\vcthreediagjkpre}{1.23}
\newcommand{\vcthreejklopre}{$-$18}
\newcommand{\vcthreejkhipre}{+28}
\newcommand{\vrecvNslab}{5889}
\newcommand{\vrecvNngc}{217365}
\newcommand{\vrecvshiftrms}{10.3}
\newcommand{\vrecvshiftmed}{8.9}
\newcommand{\vrecvdisp}{7.8}
\newcommand{\vrecvdisplos}{4.7}
\newcommand{\vrecvdrmspre}{1.00}
\newcommand{\vrecvdrmspost}{0.87}
\newcommand{\vrecvzoomone}{300}
\newcommand{\vrecvzoomtwo}{240}
\newcommand{\vrecvzoomf}{80}
\newcommand{\vrppimodeltext}{Orange dashed contours are the best-fit model with the redshift-error tail ($b\sigma_8=1.851$, $f\sigma_8=0.549$, and a fraction $p=0.180$ of clusters with a line-of-sight error $\sigma_o=78.3\hMpc$), and green dotted contours are the linear Kaiser prediction for the Planck fiducial cosmology ($f\sigma_8=0.475$ at $z=0.405$), with the same $b\sigma_8$ and no redshift errors.}
\newcommand{\vrppistretchtext}{The data are instead stretched along $\pi$ at small separations, as the model with the tail predicts: the redshift errors of the clusters whose redshift is that of a member galaxy act like Fingers of God.}
\newcommand{\vrppifittext}{The model is drawn for reference and is not a measurement of $f\sigma_8$; $\xi$ is used for the BAO position only. }
\newcommand{\vrppibodytext}{A fit of the redshift-space model to $\xi_0$ and $\xi_2$ over $\vrppismin\le s\le\vrppismax\hMpc$, jointly with $w_p(r_p)$, gives $f\sigma_8=\vrppifsmed\pm\vrppifserr$ and $b\sigma_8=\vrppibsmed\pm\vrppibserr$, with $\chi^2=\vrppichisq/\vrppindof$; $\xi_0$ and $\xi_2$ alone give $f\sigma_8=\vrppibasefsmed\pm\vrppibasefserr$, with $\chi^2=\vrppibasechisq/\vrppibasendof$. The Planck fiducial value is $f\sigma_8=\vrppifsplanck$. The same fits in the stage-1 analysis of the companion paper, with the earlier random catalog, gave $f\sigma_8=0.362\pm0.043$ with $\chi^2=82.1/65$ for the joint fit and $0.30^{+0.09}_{-0.07}$ with $\chi^2=32.1/44$ for $\xi_0$ and $\xi_2$ alone. We infer, without a direct test, that the smooth $n(z)$ given to the present random catalog in each cell raises $\xi_0$ on large scales, which would raise both $\chi^2$ and $f\sigma_8$. We therefore use $\xi$ for the BAO position only: the fitted model in Fig.~\ref{fig:rppi} is drawn for reference and is not a measurement of $f\sigma_8$. }
\newcommand{\vrppifsplanck}{0.475}
\newcommand{\vrppifsmed}{0.545}
\newcommand{\vrppifserr}{0.049}
\newcommand{\vrppibsmed}{1.854}
\newcommand{\vrppibserr}{0.026}
\newcommand{\vrppichisq}{283.2}
\newcommand{\vrppindof}{65}
\newcommand{\vrppibasefsmed}{1.334}
\newcommand{\vrppibasefserr}{0.203}
\newcommand{\vrppibasechisq}{115.0}
\newcommand{\vrppibasendof}{44}
\newcommand{\vrppismin}{30}
\newcommand{\vrppismax}{150}

\begin{document}

\title{Baryon Acoustic Oscillations from the all-sky Wen \& Han cluster catalog and the 2MASS Redshift Survey}
\author{Mart\'in G. Richarte}
\email{martin@df.uba.ar (Corresponding author)}
\affiliation{Centro de Ci\^encias Naturais e Humanas, Universidade Federal do ABC, Santo Andr\'e, SP, Brazil}
\affiliation{Departamento de F\'{\i}sica, Facultad de Ciencias Exactas y Naturales, Universidad de Buenos Aires, Buenos Aires, Argentina}
\author{Diego G. Lambas}
\email{diego.garcia.lambas@unc.edu.ar}
\affiliation{Instituto de Astronom\'{\i}a Te\'orica y Experimental (IATE), CONICET-UNC, C\'ordoba, Argentina}
\affiliation{Observatorio Astron\'omico de C\'ordoba (OAC), UNC, C\'ordoba, Argentina}

\begin{abstract}
Baryon acoustic oscillations (BAO) in the clustering of galaxies and quasars
now measure distances to precision of about one percent although galaxy clusters have so far
given only marginal signals. We measure the BAO scale in the correlation
function of \vNclus\ galaxy clusters with spectroscopic redshifts at
$0.1<z<0.7$ from the Wen and Han (2024) all-sky catalog. After reconstruction of the
density field, the BAO peak has a significance of $\vsigpost\sigma$ against a
model without it, and gives
$\DV/\rd=\vDVpost\pm\vDVpoststat\,({\rm stat})\pm\vDVpostsys\,({\rm sys})$ at an effective redshift
$z_{\rm eff}=\vzeffv$, a \vfracposttot\% distance. 
We find that the statistical uncertainty is
\vratiostat\ times smaller than that of the best earlier cluster measurement.
The distance agrees with the Dark Energy Spectroscopic Instrument (DESI) Data
Release 2 within $\vnsdesipost\sigma$ and lies $\vnsplanckpost\sigma$ below
Planck. We also measure the BAO scale in the \vTNgal\ galaxies of the 2MASS Redshift Survey with a mean effective redshift (weighted by $n\,w^2$ with FKP weights), $z_{\rm eff}=\vTzeff$
For this all-sky galaxy sample we obtain
$\aiso=\vTpreA^{+\vTpreAup}_{-\vTpreAlo}$ before reconstruction.
\end{abstract}

\maketitle

\heading{Introduction}The sound horizon at the drag epoch, $\rd\simeq147$\,Mpc, leaves a peak in the
correlation function of any tracer of the matter density. Measured across and
along the line of sight, this baryon acoustic oscillation (BAO) scale gives the
distances $\DM(z)/\rd$ and $\DHz(z)/\rd$, or their combination $\DV(z)/\rd$.
Galaxy and quasar surveys now measure it to about one
percent~\cite{Eisenstein2005,Cole2005,Alam2017,DESI2024III}.

Galaxy clusters sit in the most massive halos. Their high bias, $b\simeq2$--4,
partly makes up for their low density. The cluster correlation function was first measured in rich Abell clusters~\cite{BahcallSoneira1983} and later in the APM and SDSS cluster samples~\cite{Croft1997,Bahcall2003}, and optical cluster finders applied to SDSS and DES imaging now give catalogs of more than $10^5$ clusters~\cite{Koester2007,WenHanLiu2012,WenHan2015,Rykoff2014,Rykoff2016}. Cluster counts and weak lensing from large optical, X-ray and Sunyaev-Zeldovich surveys now constrain cosmology~\cite{PlanckSZ2015,Bocquet2019,Hilton2021,DESY1clusters2020,Ghirardini2024}, and the clustering amplitude adds constraints on cosmology and on the cluster mass scale~\cite{Mana2013,Sereno2015,Marulli2018,Lesci2022,To2021,Fumagalli2024SDSS,Fumagalli2024}. The position of the brightest cluster
galaxy (BCG) is also little affected by small-scale motions. The cluster BAO
was seen first with photometric redshifts~\cite{Estrada2009,Huetsi2010} and
then in spectroscopic samples of the Sloan Digital Sky Survey
(SDSS)~\cite{Hong2012,Veropalumbo2014,Veropalumbo2016}. The largest of these,
79\,091 clusters at $z\le0.5$, gave a signal at $3.7\sigma$ and $\DV$ to
3.8\%~\cite{Hong2016}. Shao et al.\ \cite{Shao2026} have recently shown the promise of the WH24 catalog for this purpose, using its spectroscopic clusters at $0.17\le z\le0.5$ in two redshift bins without reconstruction. No cluster sample has yet reached $4\sigma$ after reconstruction of the density field. Here we extend that sample to $0.1<z<0.7$ and apply reconstruction.

\begin{figure}[t]\centering
\includegraphics[width=\columnwidth]{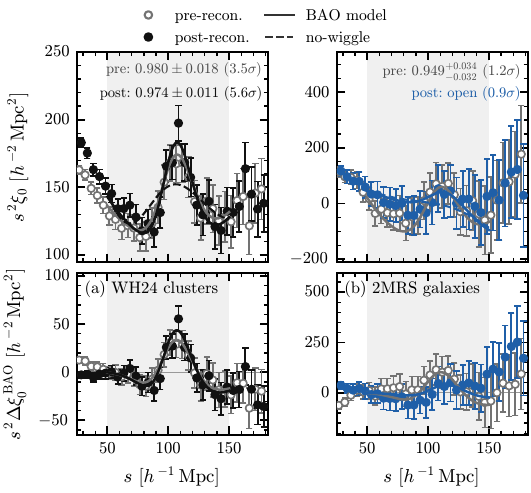}
\caption{Monopole of two samples, times $s^2$, before (open gray) and after
(filled) reconstruction, shifted by $\mp0.9\hMpc$ for clarity. (a) The
\vNclus\ \WH\ spectroscopic clusters at $0.1<z<0.7$ (black); errors from the
calibrated covariance. (b) The \vTNgal\ 2MRS galaxies at $\zeff=\vTzeff$
(blue); errors from \vTnmock\ Zel'dovich mocks. Errors are the
square root of the diagonal (no Hartlap factor); the gray band is the fitting
range, $50\le s\le150\hMpc$. Top: data with the best-fit BAO model (solid) and,
after reconstruction only, the no-wiggle model (dashed). The labels give
$\aiso$ with its error and, in brackets, the significance $\sqrt{\Delta\chi^2}$ of each fit as recorded:
(a) $\vsigpre\sigma$ before and $\vsigpost\sigma$ after reconstruction;
(b) $\vTpresig\sigma$ and $\vTpostsig\sigma$. A label marked ``open'' means that the interval $\Delta\chi^2\le1$ of $\aiso$
reaches the prior edge (\vTpriorlo\ or \vTpriorhi); this happens in (b) after reconstruction.
Such a label gives no value and no error, only the closed side of the interval
if there is one.
Bottom: the BAO feature, the data minus the smooth part of the best-fit model,
with the BAO part of the model (solid).}
\label{fig:poles}
\end{figure}

The catalog of Wen and Han~\cite{WenHan2024} (\WH) lists 1\,581\,179 clusters
found in the Legacy Imaging Surveys of the Dark Energy Spectroscopic
Instrument (DESI)~\cite{Dey2019}, over about $24\,000\,{\rm deg}^2$; 338\,841
of them have a spectroscopic redshift. Here we measure the BAO scale in the
clusters with a spectroscopic redshift, before and after reconstruction. We
then measure it in the nearby galaxies of the 2MASS Redshift Survey.
Figure~\ref{fig:poles} shows the two measurements. We
use the flat $\Lambda$CDM cosmology of the Planck 2018
TT,TE,EE+lowE+lensing+BAO chain~\cite{Planck2018}, with $h=0.6766$ and
$\Omega_{\rm m}=0.3111$, for which $\rd^{\rm fid}=\vtrd\hMpc$. All distances
are measured relative to it.

\heading{Data}The sample holds \vNclus\ clusters with a spectroscopic redshift
at $0.1<z<0.7$ over an effective area of $\vArea\,{\rm deg}^2$, at an effective
redshift $\zeff=\vzeffv$, the mean redshift of the pairs at 50--$150\hMpc$.

The random catalog must follow the selection of the clusters without copying
their clustering. Dust, stars, and the depth and seeing of the imaging change
how many member galaxies are found, and so which clusters pass the richness
cut. We therefore fit the angular density of the randoms with a Poisson
regression of the cluster counts on 20 maps that exist independently of the
catalog. The spectroscopic completeness has no
external map, so each spectroscopic sub-area of the sky has a free
normalization. Along the line of sight, a histogram of the cluster redshifts
would copy their radial clustering into the randoms~\cite{deMattia2019}. We
therefore draw random redshifts from a smooth $n(z)$ in each sky cell. The
catalog is 30 times denser than the clusters and reuses no cluster position or
redshift.

\begin{figure*}[t]\centering
\includegraphics{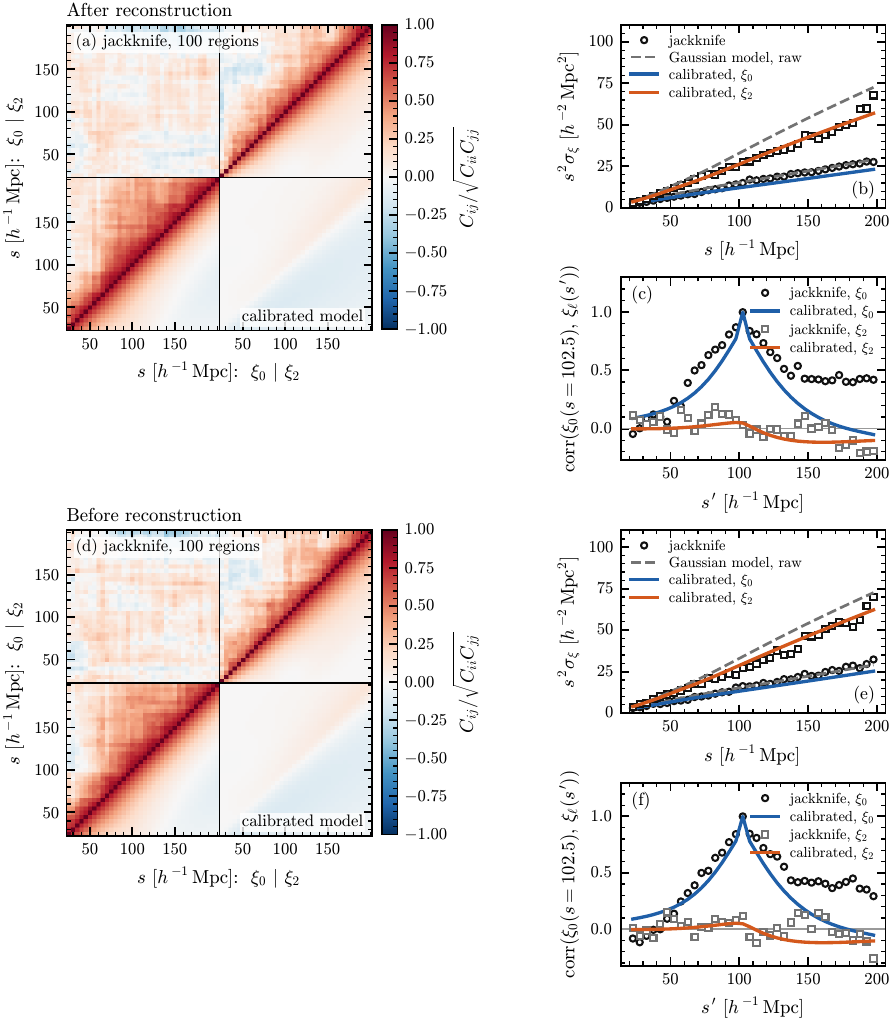}
\caption{Covariance of the monopole and quadrupole of the \vNclus\ \WH\
spectroscopic clusters for $20\le s\le200\hMpc$ in $5\hMpc$ bins, 36 bins for
each multipole, after (top, a--c) and before (bottom, d--f) reconstruction;
the fits use $50\le s\le150\hMpc$. Values in parentheses are before
reconstruction. (a, d) Correlation coefficients $C_{ij}/\sqrt{C_{ii}C_{jj}}$
from $-1$ (blue) to $+1$ (red). The black lines separate the $\xi_0$ block (bottom left) from the
$\xi_2$ block (top right) and the cross blocks. The upper-left triangle is the
jackknife over \vnjk\ regions; the lower-right triangle is the calibrated
Gaussian model used in the fits, with
volume and shot-noise factors $A=\vcthreeApost$ ($\vcthreeApre$) and
$a=\vcthreeapost$ ($\vcthreeapre$). Neighboring bins of $\xi_0$ are strongly
correlated: at $s=102.5\hMpc$ the coefficient with the next bin is
\vcthreenbrpost\ (\vcthreenbrpre) in the model and \vcthreenbrjkpost\
(\vcthreenbrjkpre) in the jackknife, and it is still \vcthreetwentyfivepost\
(\vcthreetwentyfivepre) at $25\hMpc$ separation in the model. The coefficient
between $\xi_0$ and $\xi_2$ at the same $s$ is small, \vcthreexipost\
(\vcthreexipre). The jackknife shows the same structure with the noise of its
\vnjk\ realizations; the model is smooth because it is computed from the
window and the power spectrum. (b, e) Square root of the diagonal, times
$s^2$: calibrated
model for $\xi_0$ (blue) and $\xi_2$ (orange), jackknife (open symbols) and
Gaussian model before calibration (gray dashed). The raw Gaussian errors are a
median \vcthreegausspost\ (\vcthreegausspre) times the calibrated ones. Over
the 72 bins of $\xi_0$ and $\xi_2$ the median ratio of jackknife to calibrated
errors is \vcthreejkmedpost\ (\vcthreejkmedpre), and every bin lies between
\vcthreejklopost\% and \vcthreejkhipost\% (\vcthreejklopre\% and
\vcthreejkhipre\%). In the fitting range of $\xi_0$ the jackknife variances
are \vcthreediagjkpost\ (\vcthreediagjkpre) times the calibrated ones on
average. (c, f) The row of (a, d) at $s=102.5\hMpc$, near the BAO peak: the
coefficient of $\xi_0(102.5\hMpc)$ with $\xi_0$ (blue) and $\xi_2$ (orange) at
all $s'$; points are the jackknife, lines the model.}
\label{fig:cov}
\end{figure*}

\heading{Method}We count pairs with \textsc{pycorr}~\cite{pycorr} and the
Landy-Szalay estimator~\cite{LandySzalay1993}, and measure the multipoles
$\xi_{0,2}$ in $5\hMpc$ bins. A jackknife matrix alone is noisy. We therefore
model the covariance of the clusters as Gaussian, for the varying density of
the survey, and fit its volume and shot-noise factors to a jackknife over
\vnjk\ regions~\cite{OConnell2016}, separately before and after reconstruction
(Fig.~\ref{fig:cov}).

\begin{figure*}[t]\centering
\includegraphics[width=\textwidth]{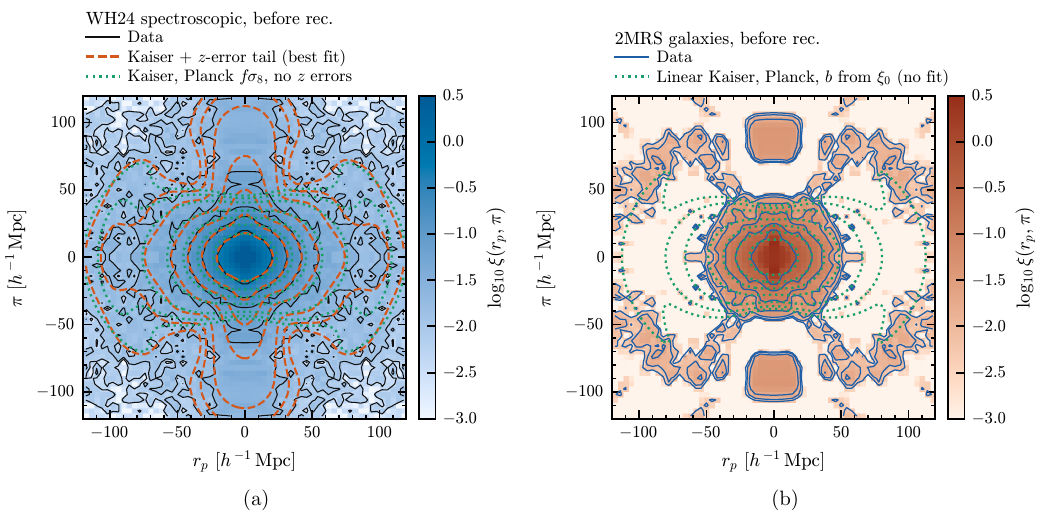}
\caption{Two-dimensional correlation function $\xi(r_p,\pi)$ before
reconstruction, in $4\hMpc$ bins, measured in one quadrant and mirrored; the
color is $\log_{10}\xi$ from $-3$ to $0.5$. (a) All \vNclus\ \WH\
spectroscopic clusters. Black contours are the data at $\xi=0.005$, 0.01, 0.02, 0.05, 0.1, 0.2 and
0.5. \vrppimodeltext\ Without redshift errors, redshift-space distortions
would flatten the contours along $\pi$. \vrppistretchtext\ \vrppifittext The BAO scale lies near
$s=\sqrt{r_p^2+\pi^2}=100\hMpc$, where the signal is too weak to show in this
map. (b) The \vTNgal\ 2MRS galaxies at $\zeff=\vTzeff$, from the pair counts,
weights and random catalog of the 2MRS correlation function of
Fig.~\ref{fig:poles}b, with the same bins and color range on a warm color
scale. Blue contours are the data at the same levels. Green dotted contours are
the linear Kaiser prediction for the Planck fiducial cosmology
($f\sigma_8=0.436$ at $z=0.048$), with the bias $b=1.72$ of the fit of $\xi_0$
at $20<s<40\hMpc$ and no velocity dispersion; no model is fitted to
$\xi(r_p,\pi)$. Across the line of sight the data follow this prediction
($\xi=1.31$ and 0.24 at $r_p=10$ and $26\hMpc$, against 1.35 and 0.23). Along
the line of sight, at $r_p<4\hMpc$ and $\pi=10\hMpc$, the data are 3.7 times
the prediction (3.04 against 0.82): the peculiar velocities of galaxies in
groups and clusters stretch the contours along $\pi$ (Fingers of God).}
\label{fig:rppi}
\end{figure*}

The template uses the linear power spectrum from \textsc{CAMB}~\cite{Lewis2000}.
Large-scale motions distort and smear the peak, so the template includes Kaiser
distortions~\cite{Kaiser1987} and BAO damping~\cite{Eisenstein2007}, with
damping scales $(\vtsperp,\vtspar)\hMpc$ across and along the line of sight
before reconstruction. It also includes the tail of the redshift errors of
clusters whose redshift comes from a member galaxy; Fig.~\ref{fig:rppi}
shows its effect on the two-dimensional correlation function. \vrppibodytext The broadband
shape of $\xi$ is uncertain, so we fit the peak position with a free amplitude
and smooth terms~\cite{Xu2012},
\begin{equation}
\xi_0^{\rm m}(s)=B\,\xi_{0,{\rm t}}(\alpha s)+a_1s^{-2}+a_2s^{-1}+a_3 ,
\label{eq:isofit}
\end{equation}
over $50\le s\le150\hMpc$, with
$\alpha\equiv\aiso=(\DV/\rd)/(\DV/\rd)^{\rm fid}$ and $B\ge0$. The joint fit
of $\xi_0$ and $\xi_2$ gives $\aperp$ and $\apar$~\cite{Xu2013}; we use only
the BAO position in $\xi_2$. The significance is $\sqrt{\Delta\chi^2}$ against
the same fit with a no-wiggle template. The systematic error
adds in quadrature the scatter $\sigma_b$ of $\alpha$ over the lognormal mocks
and the shift $\sigma_{\rm var}$ under four fit changes fixed in advance.

\heading{Results}The fit before reconstruction (Fig.~\ref{fig:poles}a,
Table~\ref{tab:bao}) gives
$\aiso=\valphapre\pm\valphaprestat\,({\rm stat})\pm\vsigsysisopre\,({\rm sys})$
with $\chi^2/{\rm dof}=\vchisqpre/15$, that is,
$\DV/\rd=\vDVpre\pm\vDVprestat\,({\rm stat})\pm\vDVpresys\,({\rm sys})$ at
$\zeff=\vzeffv$, a \vfracpretot\% distance. The no-wiggle model is worse by
$\Delta\chi^2=\vdchisqpre$, a significance of $\vsigpre\sigma$. On the
\vgtwon\ lognormal mocks, the random catalog changes this significance by
$\vdsig\pm\vdsigerr\sigma$. A random catalog without the imaging-region
amplitudes gives a consistent $\aiso$.

We test the size of the errors on \vcovn\ Zel'dovich mocks of the same
selection, fitted in the same way. With the covariance model used for the data,
the pull of $\alpha$ has a width of \vcovDpre\ before and \vcovDpost\ after
reconstruction, inside the limits fixed in advance.

\begin{figure*}[t]\centering
\includegraphics{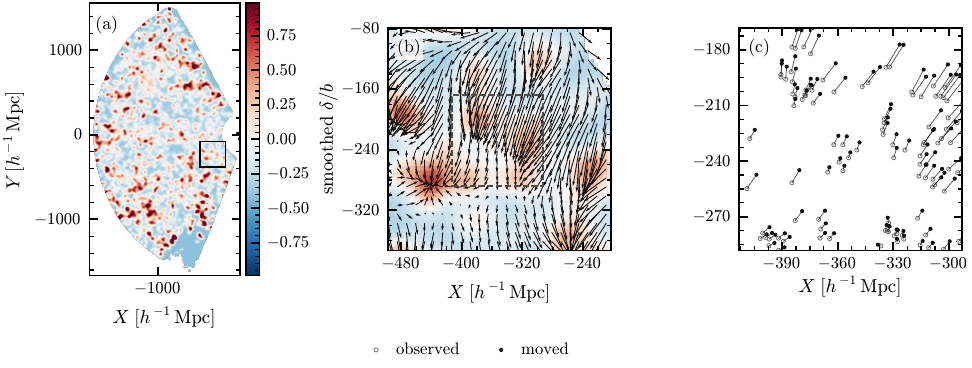}
\caption{Reconstruction of the \vrecvNngc\ \WH\ spectroscopic clusters of the
northern Galactic cap, in a slab $30\hMpc$ thick around the equatorial plane
($|Z|<15\hMpc$, \vrecvNslab\ clusters), with the observer at the origin, and
the settings of our fits (iterative Fourier method, three iterations,
$15\hMpc$ Gaussian smoothing, $b=2.47$, $f=0.724$). (a) Smoothed density
contrast divided by the bias, inside the survey; the square marks the region
of panel (b), $\vrecvzoomone\hMpc$ on a side. (b) Real-space contrast after
the three iterations, with the same color scale, and the in-plane part of the
estimated displacement, drawn five times longer than it is. Clusters flow
from the blue regions toward the red ones. (c) Clusters in the dashed square
of panel (b), at their observed positions (open gray circles) and after
reconstruction (filled black points), joined by lines. In the slab the rms
shift is $\vrecvshiftrms\hMpc$ (median $\vrecvshiftmed\hMpc$); over the whole
cap the rms displacement is $\vrecvdisp\hMpc$, $\vrecvdisplos\hMpc$ along the
line of sight.}
\label{fig:recslice}
\end{figure*}

\begin{figure*}[t]\centering
\includegraphics{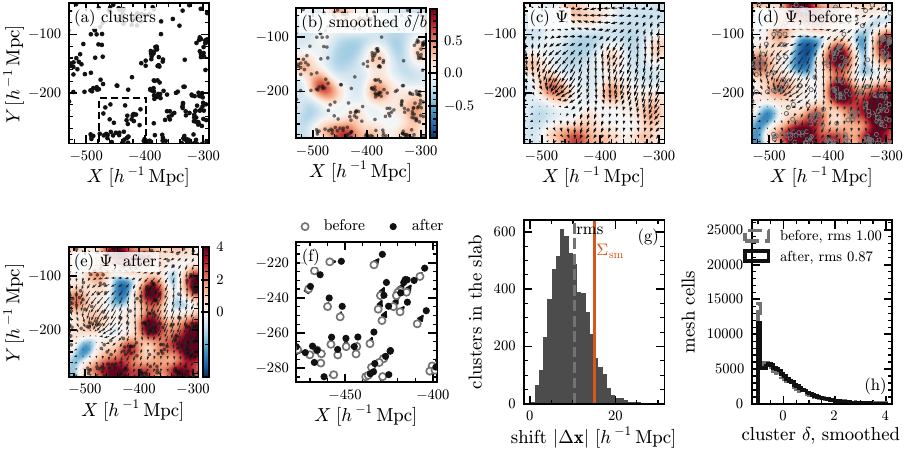}
\caption{Reconstruction step by step, in the slab $|Z|<15\hMpc$ of
Fig.~\ref{fig:recslice} (\vrecvNslab\ clusters). (a) Observed cluster
positions in a $\vrecvzoomtwo\hMpc$ square. (b) The density contrast divided
by the bias, $\delta/b$ with $b=2.47$, after Gaussian smoothing with
$\Sigma_{\rm sm}=15\hMpc$ on the $6\hMpc$ mesh. (c) The displacement
$\mathbf{\Psi}$, with $f=0.724$ and in-plane components drawn twice their
length, over the real-space density contrast after three iterations. (d), (e)
The same displacement field over the density contrast of the clusters in the
slab before (d) and after (e) reconstruction, from the cluster counts
smoothed with $\Sigma_{\rm sm}$ (no randoms); the dots are the clusters at
their positions in each case (open gray before, filled black after). (f)
Clusters in the dashed square of (a), $\vrecvzoomf\hMpc$ on a side, before
(open gray) and after (filled black) they are moved by
$-\mathbf{\Psi}-f(\mathbf{\Psi}\cdot\hat{\mathbf r})\hat{\mathbf r}$; arrows
join the two positions. (g) Distribution of the shifts of all clusters in the
slab, with the rms ($\vrecvshiftrms\hMpc$, dashed) and $\Sigma_{\rm sm}$
(solid). (h) Distribution of the smoothed cluster density of (d) and (e) over
the whole slab, with an rms of \vrecvdrmspre\ before and \vrecvdrmspost\ after
reconstruction; the spike at $-1$ is empty cells.}
\label{fig:recsteps}
\end{figure*}

\heading{Reconstruction}Bulk flows move the clusters and blur the BAO peak.
Reconstruction estimates these flows from the density field and moves the
clusters back. We use the iterative Fourier
method~\cite{Padmanabhan2012,pycorr} with the settings of the DESI Data
Release 2 (DR2) analysis~\cite{DESI2025DR2}. After reconstruction the
damping scales are $(\vdampperp,\vdamppar)\hMpc$, \vdampfac\ times those
before; this factor is fitted on the \vcovn\ Zel'dovich mocks. On \vpostn\
reconstructed lognormal mocks, the random catalog shifts $\alpha$ by
$\vpostda\pm\vpostdaerr$ and the significance by
$\vpostdsig\pm\vpostdsigerr\sigma$, both within the limits fixed in advance.
Figures~\ref{fig:recslice} and~\ref{fig:recsteps} show the reconstruction of
the clusters of the northern Galactic cap.

\begin{figure*}[t]\centering
\includegraphics{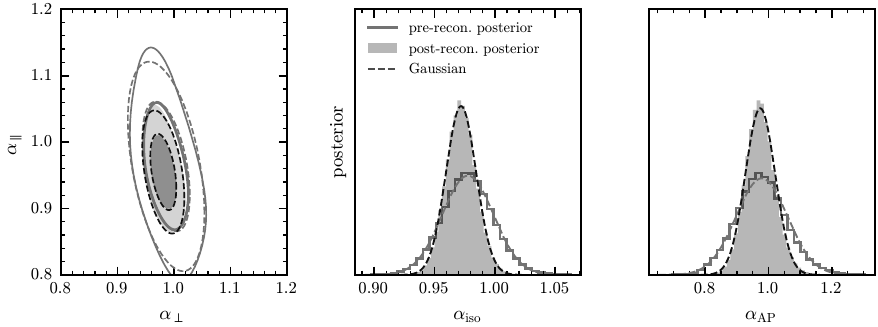}
\caption{Posteriors of the anisotropic fit of $\xi_0$ and $\xi_2$ of the \WH\
spectroscopic clusters before (gray, open) and after (black, shaded)
reconstruction, against a Gaussian with the same mean and covariance (dashed,
same color). Left: 68\% and 95\% contours of $\aperp$ and $\apar$. Middle and
right: the derived $\aiso=\aperp^{2/3}\apar^{1/3}$ and
$\alpha_{\rm AP}=\apar/\aperp$. This is the analog of Fig.~18 of
Ref.~\cite{Paillas2025}.}
\label{fig:post}
\end{figure*}

The fit after reconstruction gives
$\aiso=\valphapost\pm\valphapoststat\,({\rm stat})\pm\vsigsysisopost\,({\rm sys})$,
that is,
\begin{equation}
\DV/\rd=\vDVpost\pm\vDVpoststat\,({\rm stat})\pm\vDVpostsys\,({\rm sys})\ \text{at}\ \zeff=\vzeffv ,
\end{equation}
a \vfracposttot\% distance. Reconstruction reduces the statistical error by a
factor \vrecgain. The no-wiggle model is worse by
$\Delta\chi^2=\vdchisqpost$, a significance of $\vsigpost\sigma$. The fit has
$\chi^2/{\rm dof}=\vchisqpost/15$, a probability to exceed of about
\vptepost\ for a Gaussian likelihood. Still, on the Zel'dovich mocks the error
on $\alpha$ has the expected size after reconstruction. We quote this fit as
our result and do not combine it with the fit before reconstruction. The joint
fit of $\xi_0$ and $\xi_2$ gives $\DM/\rd=\vDMpost\pm\vDMpoststat\pm\vDMsyspost$ and
$\DHz/\rd=\vDHpost\pm\vDHpoststat\pm\vDHsyspost$ (stat, sys; posteriors in Fig.~\ref{fig:post}).

\begin{table}[t]
\caption{BAO fits of the \WH\ spectroscopic clusters. $\sigma_{\rm nom}$ is
the significance $\sqrt{\Delta\chi^2}$ against the no-wiggle model; no
significance calibrated with null simulations exists for this random catalog.
Errors are statistical, then systematic. $\aperp$ and $\apar$ are
correlated at $\vanisocorrpost$ after and $\vanisocorrpre$ before
reconstruction. The $\chi^2$ is for 15 (isotropic) and 30 (anisotropic)
degrees of freedom.}
\label{tab:bao}
\begin{ruledtabular}
\begin{tabular}{llccc}
Case & Fit & $\alpha\pm{\rm stat}\pm{\rm sys}$ & $\sigma_{\rm nom}$ & $\chi^2/{\rm dof}$ \\
\hline
After rec. & $\xi_0$, $\aiso$ & $\valphapost\pm\valphapoststat\pm\vsigsysisopost$ & \vsigposttwo & \vchisqpost/15 \\
After rec. & $\xi_0+\xi_2$, $\aperp$ & $\vaperppost\pm\vaperppoststat\pm\vsigsysperppost$ & \vanisosigpost & \vanisochisqpost/30 \\
After rec. & $\xi_0+\xi_2$, $\apar$ & $\vaparpost\pm\vaparpoststat\pm\vsigsysparpost$ & & \\
Before rec. & $\xi_0$, $\aiso$ & $\valphapre\pm\valphaprestat\pm\vsigsysisopre$ & \vsigpretwo & \vchisqpre/15 \\
Before rec. & $\xi_0+\xi_2$, $\aperp$ & $\vaperppre\pm\vaperpprestat\pm\vsigsysperppre$ & \vanisosigpre & \vanisochisqpre/30 \\
Before rec. & $\xi_0+\xi_2$, $\apar$ & $\vaparpre\pm\vaparprestat\pm\vsigsysparpre$ & & \\
\end{tabular}
\end{ruledtabular}
\end{table}

\begin{figure}[!htb]\centering
\includegraphics{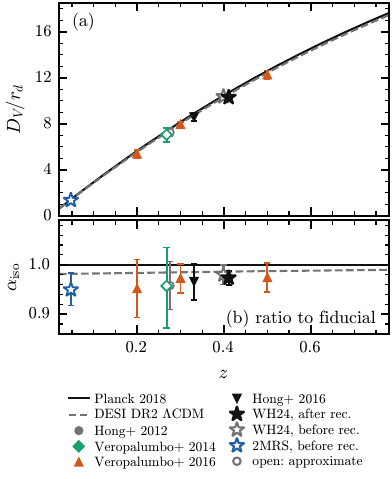}
\caption{BAO distances. (a) $\DV/\rd$ against redshift for the published
cluster BAO measurements and for this work: \WH\
clusters (stars: filled black after reconstruction, open gray before) and
2MRS galaxies before reconstruction (open blue star). After reconstruction the
2MRS interval is open and no point is drawn.
Open symbols of earlier works are approximate ($\Omega_{\rm m}h^2$ free).
Solid line: our fiducial Planck 2018 cosmology; dashed line: flat
$\Lambda$CDM fit to DESI DR2 BAO~\cite{DESI2025DR2}. (b) The same values
divided by the fiducial prediction ($\aiso$). The cluster errors are statistical and
systematic in quadrature; the 2MRS error is statistical.}
\label{fig:compare}
\end{figure}

\heading{Comparison with earlier cluster BAO}Figure~\ref{fig:compare}
collects the published cluster BAO measurements, converted to our
fiducial cosmology. Earlier spectroscopic samples reached a
distance error of 3.0\% at best~\cite{Veropalumbo2016}. Our statistical error
after reconstruction is \vratiostat\ times smaller (\vratiotot\ times with the
systematic error), and \vratiohong\ times smaller than that of the largest
earlier sample~\cite{Hong2016}. The earlier values lie at
$\alpha=0.952$--0.974, within \voffmin--$\voffmax\sigma$ of ours. All agree
within $1\sigma$ with the flat $\Lambda$CDM fit to DESI DR2
BAO~\cite{DESI2025DR2}. These samples share many clusters with ours, so the
measurements are not independent.

\begin{figure}[t]\centering
\includegraphics[width=\columnwidth]{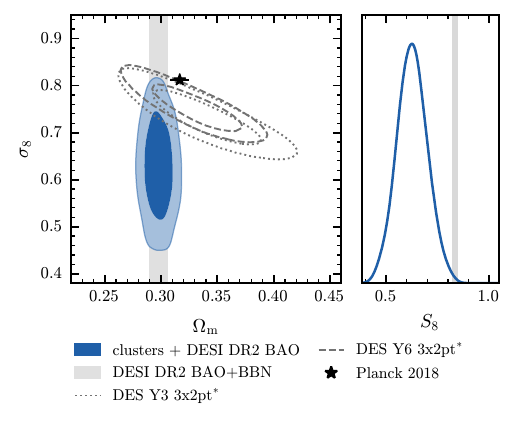}
\caption{Constraints from the spectroscopic clusters combined with DESI DR2
BAO, from the stage-1 analysis of the companion paper. (Left) 68\% and 95\% credible regions in $\Omega_{\rm m}$ and
$\sigma_8$ from $b\sigma_8$, $f\sigma_8$, the cluster BAO in two redshift
slices and the DESI DR2 BAO distances, with the normalization of the
mass-richness relation free (blue). The gray band is DESI DR2 BAO with a BBN
prior, $\Omega_{\rm m}=0.2977\pm0.0086$~\cite{DESI2025DR2}, and the star is
Planck 2018 (TT,TE,EE+lowE)~\cite{Planck2018}. The dotted and dashed lines are
DES Y3 and Y6 3$\times$2pt~\cite{DESY3_3x2pt,DESY6_3x2pt}. We do not have
their chains, so these contours ($^*$) are drawn from the published
$\Omega_{\rm m}$ and $S_8$ errors, taken as independent. (Right) Posterior of
$S_8$ for the same run; the gray band is Planck 2018.}
\label{fig:cosmo}
\end{figure}

\heading{Cosmology}Figure~\ref{fig:cosmo} and Table~\ref{tab:cosmo} show the
cosmological constraints from the clusters combined with the DESI DR2 BAO
distances. They come from the stage-1 analysis of the companion paper, which
used the earlier random catalog, and will be updated with the bias run. There,
the cluster bias $b\sigma_8$, the growth rate $f\sigma_8$ and the cluster BAO
in two redshift slices are added to the DESI DR2 BAO
distances~\cite{DESI2025DR2}, with a BBN prior and with the normalization of
the mass-richness relation free. The DESI distances fix $\Omega_{\rm m}$ and
$h$: we find $\Omega_{\rm m}=0.298\pm0.008$ and $h=0.686\pm0.006$, the DESI
values. The clusters then set $\sigma_8=0.63\pm0.07$ and $S_8=0.63\pm0.07$,
low by $1.3\sigma$ to $2.8\sigma$ against Planck 2018 and DES Y3 and Y6
3$\times$2pt~\cite{Planck2018,DESY3_3x2pt,DESY6_3x2pt}. These values are set
by the low growth rate of the clusters and remain limited by the mass scale,
which is not yet calibrated with lensing. Adding the 2MRS distance at $z=\vTzeff$, with its full
$\Delta\chi^2(\aiso)$ profile, shifts $\Omega_{\rm m}$ by $-0.17\sigma$ and $h$,
$\sigma_8$ and $S_8$ by less than $0.03\sigma$, so Fig.~\ref{fig:cosmo} is
unchanged.

\begin{table*}[t]\centering\footnotesize\setlength{\tabcolsep}{4pt}
\caption{Stage-1 results of the companion paper: our clusters combined with DESI DR2 BAO, with the mass scale free
(median and 68 percent interval), compared with DESI DR2 BAO+BBN (Table~V of
Ref.~\cite{DESI2025DR2}), Planck 2018 TT,TE,EE+lowE (Table~2 of
Ref.~\cite{Planck2018}), and DES Y3 and Y6 3$\times$2pt in $\Lambda$CDM (Eq.~17
of Ref.~\cite{DESY3_3x2pt} and Eq.~21 of Ref.~\cite{DESY6_3x2pt}). The number
in parentheses after each published value is our value minus the published
one, in units of the two errors added in quadrature. A dash marks a parameter
that is not constrained or not quoted.}
\label{tab:cosmo}
\fitwidth{\begin{tabular}{lccccc}\toprule
 & This work & DESI DR2 & Planck 2018 & DES Y3 & DES Y6 \\\midrule
$\Omega_{\rm m}$ & $0.298\pm0.008$ & $0.2977\pm0.0086$ ($0.0\sigma$) & $0.3166\pm0.0084$ ($-1.6\sigma$) & $0.339^{+0.032}_{-0.031}$ ($-1.2\sigma$) & $0.333^{+0.023}_{-0.028}$ ($-1.4\sigma$) \\
$h$ & $0.686\pm0.006$ & $0.6851\pm0.0058$ ($+0.1\sigma$) & $0.6727\pm0.0060$ ($+1.5\sigma$) & -- & -- \\
$\sigma_8$ & $0.63\pm0.07$ & -- & $0.8120\pm0.0073$ ($-2.5\sigma$) & $0.733^{+0.039}_{-0.049}$ ($-1.3\sigma$) & $0.751^{+0.034}_{-0.036}$ ($-1.5\sigma$) \\
$S_8$ & $0.63\pm0.07$ & -- & $0.834\pm0.016$ ($-2.8\sigma$) & $0.776\pm0.017$ ($-2.0\sigma$) & $0.789\pm0.012$ ($-2.2\sigma$) \\
\bottomrule
\end{tabular}}
\end{table*}

\heading{The local galaxies}Combined with the cosmic microwave background (CMB)
and supernovae, DESI BAO favor a dark energy that evolves with time over a
constant density, at $2.8$--$4.2\sigma$~\cite{DESI2025DR2}. Measurements in
the local Universe could test the late-time expansion directly. Cosmic
variance dominates their errors, and only full-sky samples can reduce it. We
therefore apply the same estimator to the \vTNgal\ galaxies of the 2MASS
Redshift Survey (2MRS; 2MASS is the Two Micron All Sky Survey)~\cite{Huchra2012}
at $\zeff=\vTzeff$, with a covariance from \vTnmock\ Zel'dovich mocks. The likelihood is far from Gaussian, so we quote the interval
where $\Delta\chi^2\le1$ of the profile likelihood (option A), a choice fixed
on the mocks before the data were fitted. Before reconstruction
$\aiso=\vTpreA^{+\vTpreAup}_{-\vTpreAlo}$, with a significance of
$\vTpresig\sigma$ (Fig.~\ref{fig:poles}b). The two-dimensional correlation
function of the galaxies (Fig.~\ref{fig:rppi}b) follows the linear Kaiser
prediction across the line of sight, with the bias $b=1.72$ of $\xi_0$. After reconstruction the interval
fills the prior. Still, this is what the mocks expect: \vTedgepre\ of the
\vTnmock\ mocks, all with a BAO peak, also reach the prior edge before
reconstruction.

\heading{Discussion and conclusions}
The all-sky \vNclus\ spectroscopic \WH\ clusters show the BAO peak at
$\vsigpost\sigma$ after reconstruction and measure $\DV/\rd$ to
\vfracposttot\% at   $z_{\rm eff}=\vzeffv$. With statistical and systematic errors, the
distance lies $\vnsdesipost\sigma$ below the flat $\Lambda$CDM fit to DESI DR2
BAO~\cite{DESI2025DR2}, \vDVdesi, and $\vnsplanckpost\sigma$ below Planck,
\vDVplanck. Before reconstruction these offsets are $\vnsdesipre\sigma$ and
$\vnsplanckpre\sigma$. Most of the cluster redshifts are BCG redshifts from
SDSS and its Baryon Oscillation Spectroscopic Survey (BOSS), so the
measurement is not independent of BOSS. Still, it adds a BAO distance from a
halo-selected tracer, whose bias would also calibrate the cluster masses.

At  $z_{\rm eff}=\vTzeff$, the 2MRS galaxies in the two galactic hemispheres give $\aiso=\vTpreA^{+\vTpreAup}_{-\vTpreAlo}$
before reconstruction, within $0.7\sigma$ of the clusters at $z=\vzeffv$, although
the peak itself is seen at only $\vTpresig\sigma$. In this small volume, cosmic
variance sets the associated uncertainty: the \vTnmock\ mocks of the same volume, all with a BAO
peak, give open intervals in \vTedgepre\ cases before and 54 after reconstruction. The
open interval of the data after reconstruction is therefore a common outcome,
not a failure of the method.

DESI spectra of the BCGs that now have only photometric
redshifts~\cite{DESI2024III} would give precise redshifts for all \WH\
clusters and would measure the redshift-error tail directly. Galaxy clusters now give a BAO distance at the percent level.

\begin{acknowledgments}
\heading{Acknowledgments}M.G.R.\ is partially supported by UFABC (Brazil) and WFSA. This work was partially supported by the Agencia Nacional de Promoci\'on Cient\'{\i}fica y Tecnol\'ogica (PICT 2015-3098, PICT 2016-1975), the Consejo Nacional de Investigaciones Cient\'{\i}ficas y T\'ecnicas (CONICET, Argentina) and the Secretar\'{\i}a de Ciencia y Tecnolog\'{\i}a de la Universidad Nacional de C\'ordoba (SeCyT-UNC, Argentina). The 2MRS data used in this work are public~\cite{Huchra2012}.
\end{acknowledgments}

\heading{Data availability}The data used in this work are public: the \WH\ cluster
catalog~\cite{WenHan2024}, the DESI Data Release 1 (DR1) and the 2MRS
catalog~\cite{Huchra2012}. The data
products that support the findings of this article are available from the
authors upon reasonable request. The code
can be shared under agreement after this article is published.

\end{document}